# Exciton Radiative Lifetime in Transition Metal Dichalcogenide Monolayers

C. Robert[1], D. Lagarde[1], F. Cadiz[1], G. Wang[1], B. Lassagne[1], T. Amand[1], A. Balocchi[1], P. Renucci[1], S. Tongay[2], B. Urbaszek[1] and X. Marie[1]

[1] *Université de Toulouse, INSA-CNRS-UPS, LPCNO, 135 Ave. de Rangueil, 31077 Toulouse, France*
[2] *School for Engineering of Matter, Transport and Energy, Arizona State University, Tempe, Arizona 85287, USA*

*We have investigated the exciton dynamics in transition metal dichalcogenide mono-layers using time-resolved photoluminescence experiments performed with optimized time-resolution. For $MoSe_2$ monolayer, we measure $\tau_{rad}^0$ = 1.8 ± 0.2 ps that we interpret as the intrinsic radiative recombination time. Similar values are found for $WSe_2$ mono-layers. Our detailed analysis suggests the following scenario: at low temperature ($T \lesssim 50$ K), the exciton oscillator strength is so large that the entire light can be emitted before the time required for the establishment of a thermalized exciton distribution. For higher lattice temperatures, the photoluminescence dynamics is characterized by two regimes with very different characteristic times. First the PL intensity drops drastically with a decay time in the range of the picosecond driven by the escape of excitons from the radiative window due to exciton-phonon interactions. Following this first non-thermal regime, a thermalized exciton population is established gradually yielding longer photoluminescence decay times in the nanosecond range. Both the exciton effective radiative recombination and non-radiative recombination channels including exciton-exciton annihilation control the latter. Finally the temperature dependence of the measured exciton and trion dynamics indicates that the two populations are not in thermodynamical equilibrium.*

## I. Introduction :

Transition-metal dichalcogenides (TMDC) such as $MX_2$ (M=Mo, W ; X=S, Se, Te) are a new exciting class of atomically flat, two-dimensional materials for electronics and optoelectronics. In contrast to graphene, monolayer $MX_2$ have a direct band gap yielding interesting luminescence properties in the visible region of the optical spectrum[1,2]. This has been exploited to fabricate light-emitting diodes and laser prototypes using TMDC monolayer as active region[3,4,5,6,7]. The combined presence of inversion symmetry breaking and strong spin-orbit coupling in these monolayers (ML) also yields very original spin/valley properties, which are usually probed by optical spectroscopy techniques[8,9,10,11,12].

Recent experimental and theoretical studies also demonstrated that the optical properties of ML TMDC are governed by strongly bound excitons with binding energy of the order of 500 meV[13,14,15,16,17,18,19]. This is very promising for possible applications based on strong light-matter coupling[20,21]. Therefore the knowledge of the exciton lifetime in these new 2D semiconductors is crucial from both a fundamental and applied physics point of view.

The exciton and carrier dynamics in TMDC monolayers has been investigated by pump-probe absorption or reflectivity experiments[22,23,24,25,26,27]. Thanks to very good time-resolution (~ 100 fs), these experiments evidenced complex dynamics usually described by multi-exponential decay times[28]. However the identification of radiative recombination times among the multiple mechanisms of collective electronic excitations in the transient absorption or reflectivity spectra is usually challenging[26]. The ideal spectroscopy tool to investigate the radiative recombination properties is time-resolved photoluminescence (TRPL). Nevertheless it is usually characterized by a modest time-resolution (~ tens of ps) that can prevent the observation of key features associated with fast radiative recombination times[29,30,31]. At low temperature ($T \lesssim 10$ K) , recent TRPL measurements in $MoS_2$, $MoSe_2$ and $WSe_2$ ML evidenced a very fast emission decay time of ~ 3–5 ps limited by the time resolution of the set-up[32,33,34,35]. Though exciton radiative recombination times in the range of 0.1–1 ps have been calculated[36,37], a non-ambiguous measurement of the exciton radiative recombination in TMDC monolayers is still lacking.

Here we present a comprehensive investigation of the exciton dynamics using high resolution time-resolved photoluminescence technique. These experiments reveal the transient evolution from a non-thermal regime where the population decay is dominated by the intrinsic exciton radiative recombination to a thermalized exciton distribution where both radiative and non-radiative recombination channels have to be considered. As a model system, we present a detailed investigation of the $MoSe_2$ ML from $T = 7$ to 300 K, where the neutral and charged exciton PL signals are not obscured by defects-related emission. Measurements on $WSe_2$ ML are also discussed. In order to draw some general conclusions we have investigated $MoSe_2$ ML exfoliated from different bulk materials and with different environment (suspended or directly deposited on $SiO_2$/Si substrate). This allows us to distinguish between intrinsic or extrinsic recombination mechanisms.

Thanks to an optimized time-resolution of the set-up based on a sub-ps Streak Camera and the quasi-resonant excitation of the exciton ground state we managed to resolve for the first time the intrinsic exciton radiative recombination time which dominates the exciton decay in the temperature range $T = 7$–50 K. We find $\tau_{rad}^0 = 1.8 \pm 0.2$ ps in $MoSe_2$ ML for $T = 7$ K. For larger temperatures, the initial decay time decreases to sub-ps values as a consequence of the escape time of exciton out of the radiative window due to exciton-phonon interactions. Longer recombination times (~ ns) are measured after this initial transient when thermalized excitons can be considered. This longer decay time is controlled by the effective exciton radiative recombination time, the interplay between bright and dark fine-structure excitons[30,38,39,40] and the non-radiative recombination channels including the exciton-exciton annihilation rate[29,41]. Finally the simultaneous measurement of the exciton and trion dynamics as a function of temperature or resident carrier concentrations lead us to the conclusion that the two populations are not in thermodynamical equilibrium in contrast to recent assumptions[30,42].

The paper is organized as follows. The next section presents the samples characteristics and the experimental setup. Then we discuss the theoretical background of the radiative recombination of Wannier-Mott excitons in 2D semiconductors (Section III). The measurements of the very fast PL decay time associated with the exciton intrinsic radiative recombination times are presented in section IV. The dependence of the thermalized excitons kinetics as a function of temperature is discussed in section V with all the radiative and non-radiative mechanisms. Finally the trion radiative recombination time is measured in section VI. The drastic decrease of trion photoluminescence decay time with temperature and its dependence on the resident carrier density are discussed. Finally we discuss in section VII several issues related to the exciton formation process, the free versus localized exciton dynamics and the key role played by the different non-optically active (dark) exciton states.

## II. Samples and experimental set-ups:

$MoSe_2$ and $WSe_2$ ML flakes are obtained by micro-mechanical cleavage of a bulk crystals on $SiO_2$/Si substrates using viscoelastic stamping[35,43]. The ML region is identified by optical contrast and very clearly in PL spectroscopy. We present in this paper the results obtained on four different ML flakes: ML A and ML B are obtained with MoSe$_2$ bulk materials from 2D semiconductors company and Arizona State University respectively. The latter was grown using low-pressure vapour transport (LPVT) technique to achieve high optical quality materials[35,38]. MoSe$_2$ ML C is suspended between two Cr/Au (5/50 nm) electrodes pre-patterned by E-beam lithography on SiO$_2$(90 nm)/Si substrate. The distance between the two electrodes is 4 μm (*i.e.* larger than the laser spot). The application of a bias voltage between the top Cr/Au electrodes and the p-doped Si substrate (used as back gate) enables us to tune electrically the resident carrier density. For comparison, we also investigated a WSe$_2$ ML flake (ML D) obtained from bulk material from 2D semiconductors and transferred onto a simple SiO$_2$/Si substrate. A standard micro-PL set-up is used to record the emission dynamics in the temperature range $T = 7$–300 K. The laser excitation and PL detection spot diameters are ≈ 2 μm, *i.e.* smaller than the ML flake diameter. It is also larger than the estimated exciton diffusion length in TMDC ML[24]. For time-resolved photoluminescence experiments, the flakes are excited by ~ 150 fs pulses generated by a tunable mode-locked Ti:Sa laser with a repetition rate of 80 MHz. For MoSe$_2$ MLs similar results have been obtained for laser excitation energy 1.779 eV or 1.746 eV. For WSe$_2$ ML, the excitation energy was set to 1.797 eV. The laser average power on the flakes lies in the linear response regime with a typical value of 100 μW. This corresponds to a typical photogenerated exciton density $N_0 \sim 10^{11}$–$10^{12}$ cm$^{-2}$ assuming an absorption of about 1% for the considered excitation wavelengths as roughly estimated from absorption and excitation spectra[43,44]. In all the experiments the excitation laser is linearly polarized. The time-resolution of the detection system has been carefully optimized: the PL signal is dispersed by a double spectrometer operating in subtractive mode and detected by a synchro-scan Streak Camera C10910 with a nominal 900 fs time-resolution. By measuring the backscattered laser pulse from the sample surface we obtain the overall instrumental response of the time-resolved set-up: we find in figure 1b a characteristic instrument response of 0.8 ps (Half Width at Half Maximum) obtained by a Gaussian fit. For cw experiments, the monolayers are excited with a He-Ne laser and the PL emission is dispersed in a spectrometer and detected with a cooled Si-CCD camera. Figure 1a displays the cw PL spectrum of the MoSe$_2$ monolayer (ML A) at $T = 4$ K. Both exciton and trion (charged exciton) peaks are clearly observed as already evidenced in previous works[39,42].

### III. Exciton radiative lifetime: theoretical background

The intrinsic radiative decay of a free Wannier-Mott exciton in 2D semiconductors is due to coupling with a continuum of photon states. For light propagating perpendicular to the 2D layer, this intrinsic radiative decay can be calculated assuming conservation of the in-plane wavevector $k$. In a simple approach, the exciton intrinsic radiative decay time writes[45,46]:

$$\tau_{rad}^0 = \frac{1}{2\Gamma_0} = \frac{\hbar \varepsilon}{2 k_0} \left( \frac{E_{X^0}}{e \hbar v} \right)^2 \left( a_B^{2D} \right)^2 , \qquad (1)$$

Where $\Gamma_0$ is the radiative decay rate, $k_0 = E_{X^0} \sqrt{\varepsilon} / (\hbar c)$ is the light wave-vector in the sample ($c$ is the speed of light, $\varepsilon$ the dielectric constant and $E_{X^0}$ is the exciton transition energy; $v$ is the Kane velocity related to the inter-band matrix element of the electron momentum). The 2D exciton Bohr radius is $a_B^{2D} = \left( \varepsilon \hbar^2 \right) / \left( 4 \mu e^2 \right)$ with $\mu$ the exciton reduced mass defined as $1/\mu = 1/m_e + 1/m_h$, $m_e$ ($m_h$) are the effective mass for electron (hole) in the K valleys.

For times longer than typical exciton-exciton and exciton-phonon interaction times, a thermalized exciton population can be considered. Then the decay rate of the exciton photoluminescence is given by the thermal average of the exciton decay rate. It will depend on the thermal energy $k_B T$ and the kinetic energy of the excitons which decay radiatively (in the so called optical window). The latter is at most $E_0 = \left(\hbar k_0\right)^2 / 2M$, where $M = m_e + m_h$ is the exciton mass (see figure 4a). Using $M = 1.0\ m_0$ obtained from DFT-G$_0$W$_0$ calculations[43] ($m_e = 0.49\ m_0$ and $m_h = 0.52\ m_0$, $m_0$ is the free electron mass) and $n$ the refractive $n = \sqrt{\varepsilon} \sim 2.2$, we find $E_0 \sim 10\ \mu\text{eV}$.

Only the small fraction of exciton which occupy the states with $k < k_0$ (*i.e.* with kinetic energy smaller than $E_0$) can decay radiatively. As a consequence, the effective exciton decay time at a given temperature writes[45]:

$$\tau_{rad}^{eff} = \frac{3}{2}\frac{k_B T}{E_0}\tau_{rad}^0 \qquad (2)$$

We emphasize that this simple expression is valid for $k_B T \gg \gamma_h$, $E_0$, *i.e.* a few tens of Kelvin in our case ($\gamma_h$ is the exciton homogeneous linewidth[47]). Equation (2) also considers an averaging over all the fine structure (bright and dark) exciton spin states. As a consequence it should apply when $k_B T$ is also of order of the bright-dark exciton splitting[48,49,50].

At high lattice temperature, the PL decay time is usually not only dominated by pure radiative recombination but also by non-radiative recombination channels. The time evolution of the exciton population with density $N$ can be described by the simple rate equation:

$$\frac{dN}{dt} = -\frac{N}{\tau_{rad}^{eff}} - \frac{N}{\tau_{nr}} - \gamma\frac{N^2}{2} \qquad (3)$$

where $\tau_{nr}$ is the non-radiative recombination time corresponding to capture time on defects and $\gamma$ the non-radiative exciton-exciton annihilation rate[29,41,5152]. It has been shown recently that the latter process (Auger type) is dominant at room temperature[53,54].

## IV. Exciton intrinsic radiative lifetime

Figure 1b displays the photoluminescence dynamics at $T = 7$ K for the three different MoSe$_2$ monolayer samples (ML A-C). Following a rise time smaller than 1 ps (which cannot be resolved), we observe a mono-exponential decay time on more than two orders of magnitude with a characteristic value $\tau_{rad}^0 = 1.8 \pm 0.2$ ps. In contrast to previous TRPL measurements limited by a time-resolution of about 3–5 ps[34,35], the PL decay time measured here is clearly longer than the measured instrumental response (see the "laser" curve in figure 1b for comparison). A deconvolution procedure using the measured time-resolution of the set-up does not change significantly the $\tau_{rad}^0$ value. Remarkably figure 1b shows that the PL decay time is similar for the three samples corresponding to monolayers exfoliated from different MoSe$_2$ bulk materials or with different dielectric environments (suspended or not). This is a strong indication that this decay time corresponds to an intrinsic radiative recombination mechanism and not to non-radiative recombination on defects which are expected to vary from one material to another one. Moreover we have measured very similar PL decay times for WSe$_2$ monolayer (see figure 2); a mono-exponential fit yields $\tau^0_{rad} \sim 2 \pm 0.2$ ps. Though this low temperature PL decay time is likely governed by the exciton intrinsic radiative recombination time, we should also consider the possible interplay between the bright and dark excitons (composed of parallel and anti-parallel conduction and valence band electron spins respectively). It has been shown recently that the dark states lie at lower (higher) energy compared to the bright ones in WSe$_2$ (MoSe$_2$) mono-layers with typical splitting energy of 10–30 meV [30,38,39,40,50]. For the sake of simplicity, the detailed investigation presented below will be focused on MoSe$_2$ mono-layers where we can consider that the transfer from bright (at low energy) to dark states (at high energy) is negligible at low temperatures [50].

Let us emphasize that the laser excitation energy is far below the free carriers band gap: it lies between the 1s and 2s exciton states [43]. The laser excitation energy used here for the MoSe$_2$ mono-layers (1.746 eV) corresponds to a difference between excitation and exciton detection energies of about 90 meV. Considering the large spectral width of the 150 fs laser (~ 10 meV), this allows the very fast exciton photo-generation near $k \sim 0$ via a three LO phonon emission process[55,56]. Recent resonant Raman experiments in TMDC mono-layers demonstrated clearly the strong efficiency of phonon interaction[57,58].

As a consequence the quasi-resonant laser excitation results in the photo-generation of a non-thermal exciton distribution close to $k = 0$ which couples very efficiently to light as all the excitons lie in the radiative window characterized by the energy $E_0$. This leads to a very efficient and short emission process corresponding to the intrinsic exciton radiative lifetime. The very short intrinsic radiative recombination time (~ 10 times shorter than the one measured in GaAs or CdTe quantum wells[59,60,61,62]) is the result of the strong oscillator strength associated with the very robust exciton in TMDC ML.

This interpretation is further supported by two arguments:

(i) At $T = 7$ K, all the light is emitted before $t \lesssim 10$ ps. In agreement with previous time-resolved PL measurements in MoS$_2$, MoSe$_2$ or WSe$_2$ ML with lower time-resolution[32,34,35], there is no evidence of a second slower decay time following the fast initial one. As observed for higher temperatures (see section V), this would be the fingerprint of a significant fraction of thermalized excitons populating large $k$ wave-vectors which do not couple to light (figure 4b).

(ii) The non-radiative recombination times, clearly evidenced for $T \gtrsim 100$ K (see section V) occur on a much longer time scale (hundreds of ps). As a consequence we propose that the ~ 1.8 ps decay time measured in figure 1b has an intrinsic origin.

The measured exciton decay time can be compared with the calculated intrinsic radiative recombination time of ideal 2D excitons (equation 1). Using an exciton binding energy of ~ 500 meV as measured recently[16,43], the calculated reduced exciton mass $\mu = 0.25\ m_0$ and the dielectric constant $\varepsilon = 5$ (ref. 43), equation (1) yields a calculated radiative lifetime of $\tau^0_{rad} \sim 0.3\,\text{ps}$, six times shorter than the measured one in figure 1b. For this first approximation we have used here the Kane velocity v estimated from a two-bands model as $v = \sqrt{E_g/(2m_e)}$, where $E_g$ is the free carriers band gap[43] (we assume here $m_e = m_h \sim 0.5 m_0$ ). Note that similar sub-ps intrinsic exciton radiative lifetimes values have been obtained with more sophisticated theoretical approaches[36,37]. However these calculated values are based on (i) the assumption of an ideal 2D Wannier-Mott exciton which is questionable for a 2D material based on TMDC and (ii) several parameters with very large uncertainties including the electron effective mass $m_e$ (which has not been measured yet to the best of our knowledge) and the dielectric constants with complex screening and anti-screening effects[63].

Our measurements can be well interpreted if we assume that the exciton thermalization process requires a characteristic time longer than $\tau_{rad}^0$ for a lattice temperature $T = 7$ K. This thermalization process involves both exciton-exciton and exciton-phonon interactions. For a lattice temperature $T \lesssim 40$ K, the measured initial PL decay time displayed in figure 3 does not depend on temperature: in this regime the exciton-phonon interaction time is typically longer than $\tau_{rad}^0$. This absence of variation of the PL decay time in the temperature range 7-40 K (inset of figure 3) supports again our interpretation based on a non-thermal exciton distribution. A thermal exciton population should lead to a linear increase of the exciton radiative lifetime with temperature as considered in the low temperature calculations of ref. 36 and 37 (see also equation (2)).

For larger temperatures ($T \gtrsim 50$ K), we observe that the initial PL decay time decreases. We measured the same temperature dependences for samples ML B and ML C. We emphasize that we did not observe in this temperature range any blue shift of the PL peak which could have been the fingerprint of a transient change from a localized exciton regime to a free exciton regime[64,65].

In a very simple approach, the PL decay time in this temperature range can be written as $1/\tau_{PL} = 1/\tau_{rad}^0 + 1/\tau_{escape}$, where $\tau_{escape}$ is the escape time of excitons from the radiative window driven by the exciton-phonon interaction (figure 4a). When the exciton-phonon interaction time becomes shorter than the intrinsic radiative exciton lifetime $\tau_{rad}^0$, the initial PL decay time is no more driven by $\tau_{rad}^0$. For temperatures larger than 125 K, figure 3 shows this interaction time becomes so short ($< 1$ ps) that it cannot be resolved any more. The inset in figure 3 shows the dependence of $\tau_{PL}$ as a function of temperature. From the measured dependence above 50 K, it should be possible in principle to extract the efficiency of exciton-phonon interactions as recently obtained from 2D Fourier transform spectroscopy in WSe$_2$ monolayers[47]. In this low temperature range, only absorption of phonons has to be considered and the scattering rate of $k \sim 0$ excitons outside the radiative window can be described by a linear variation with $T$ [60]. However the limited time-resolution in the present TRPL measurements prevent us to extract very accurate values of this exciton-phonon scattering efficiency. Non-linear techniques such as Four Wave mixing experiments in high quality samples will be highly desirable to extract this fundamental parameter[66].

After a few tens of ps, these fast exciton-phonon interactions will yield the establishment of a thermalized Boltzmann distribution of exciton and as a consequence longer radiative recombination times[45,60,67].

## V. Exciton effective radiative lifetime

For a thermalized exciton population at a given temperature larger than a few tens of Kelvin, the fraction of exciton which lies in the radiative window is small yielding a radiative decay time longer than the intrinsic recombination time (equation (2) and figure 4b). In the temperature range 100–300 K, we clearly observe in figure 5b "long" decay times of the order of nanosecond.

For intermediate temperatures (see for instance figure 5a at $T \sim 125$ K), one can observe in the luminescence dynamics the transition from a non-thermal regime where the fast ($\sim 1$ ps) initial decay time probes the fast escape of excitons from the radiative window (discussed in section IV) to a much slower regime with a decay time in the nanosecond range observed at long delay corresponding to the effective radiative lifetime of thermalized excitons[32]. For lattice temperatures above 200 K, figure 5b shows that the escape time is so fast that the fast decay even disappears. Assuming that the exciton intrinsic radiative lifetime corresponds to the fast PL decay time measured at $T = 7$ K ($\tau_{rad}^0 = 1.8$ ps), we get a rough estimate of the exciton effective radiative decay time at $T = 125$ K from equation (2) : $\tau_{rad}^{eff} \sim 2500$ ps . For $MoSe_2$ ML we used the following parameters: exciton mass $M = 1.0 m_0$ and refractive index $n \sim 2.2$ [17,43] . This calculated value is in good agreement with the measured long PL decay time measured in figure 5a.

Although the experimental results are consistent with this exciton effective radiative recombination scenario, a more quantitative analysis cannot be conducted since non-radiative recombination channels have to be considered for temperatures larger than ~ 100 K. The inset in figure 5b displays the variation of the cw-integrated exciton intensity as a function of temperature. For $7 < T \lesssim 70$ K, the luminescence yield is rather constant: there is no signature of non-radiative recombination channels in agreement with the results presented in figures 1-3. In this regime the exciton lifetime is mainly controlled by the very fast exciton intrinsic radiative recombination time $\tau_{rad}^0$ . For larger temperatures $T \gtrsim 100$ K, the luminescence yield decreases (inset of figure 5b) as a significant fraction of excitons lies above the radiative window, their lifetime is much longer and they start to be sensitive to non-radiative recombination channels. Thus the exciton PL decay time measured in figure 5b for $t \gtrsim 100$ ps (in the thermalized exciton population regime) is controlled both by the exciton effective radiative recombination time $\tau_{rad}^{eff}$ and the non-radiative recombination processes (see equation (3)).

Figure 5b shows that $\tau_{PL}$ decreases with the temperature for $T > 100$ K whereas the exciton effective radiative recombination time is expected to increase linearly with T (equation (2)). Note that we measured the same temperature dependences for ML B obtained with a different bulk $MoSe_2$ material and for ML C which is suspended. We conclude that for $T \gtrsim 100$ K, the main recombination channel for the thermalized exciton population is non-radiative. Two non-radiative recombination processes can be considered: the standard trapping of photogenerated excitons on defects which can be thermally activated and the exciton-exciton annihilation process (Auger type)[29,41]. It has been shown recently[53,68] that the latter is the dominant exciton recombination process at room temperature for photogenerated excitons densities larger than $10^9$ cm$^{-2}$. The photoluminescence dynamics presented in figure 5 are in agreement with this interpretation. The bold-dashed line in figure 5b at $T = 300$ K is a fit according to the solution of equation (3): $N(t) = N_0 e^{-t/\tau} / \left[1 + \gamma N_0 (1 - e^{-t/\tau}) / 2\right]$ . Using the exciton-exciton annihilation rate $\gamma = 0.35$ cm$^2$/s recently estimated in $MoSe_2$ ML[29,41], we get a very good agreement with the measured kinetics using the parameters $N_0 = 10^{11}$ cm$^{-2}$ and $1/\tau_{rad}^{eff} + 1/\tau_{nr} = 1$ ns$^{-1}$. For all temperatures larger than 100 K, the PL decay time is not mono-exponential and it has a marked ~ $1/N$ dependence as it is expected for the exciton-exciton annihilation process (see equation (3)). We observe that the efficiency of this process increases when the temperature increases as a consequence of the increased mobility of excitons[29,68]. At very low temperature ($T \lesssim 50$ K), the exciton diffusion constant is so small that the exciton-exciton annihilation process efficiency vanishes. Finally we note that equation (2) yields a calculated exciton effective radiative lifetime of about 6 ns at room temperature, in good agreement with the recently measured PL decay times in a chemically treated $MoS_2$ monolayer where non-radiative recombination channels have been suppressed[53,54].

## VI. Exciton and trion radiative lifetimes:

It is well known that the TMDC mono-layers have a significant residual doping density[42]. As a consequence, the PL spectra at low temperature exhibit a clear peak associated with the trion (charged exciton) recombination in addition to the already discussed neutral exciton $X^0$ peak (see figure 1a). In agreement with previous measurements performed in $WSe_2$ monolayers[34], we did not find in our experiments any evidence of electronic transfer from excitons to trions in $MoSe_2$ ML. For instance, the rise-time of the trion luminescence at $T = 7$ K does not correspond to the decay time of the neutral exciton (the improved time-resolution used in this work should allow us to easily resolve it). This means that the trion formation time from neutral excitons is longer than the very fast decay of the neutral exciton with the exciton intrinsic radiative recombination time $\tau_{rad}^0 \sim 1.8$ ps (figure 1b). A recent investigation based on two-colour pump-probe measurements estimated a trion formation time at low temperature of the order of 2 ps in $MoSe_2$ ML[69]. Although this result does not contradict our interpretation, the comparison has to be done with caution because of the different resident carriers densities in the two experiments.

Here the excitons and trions are photogenerated quasi-resonantly on a typical scale smaller than 1 ps following 3 and 4 LO phonon emission processes respectively. Note that the exciton-trion energy separation (~ trion binding energy) coincides with the LO phonon energy[58,70]. Remarkably, our experiments indicate that the exciton and trion population decay independently.

At $T = 7$K, we measure in figure 6a a trion PL decay time of ~ 15 ps, in agreement with previous reports on $WSe_2$ or $MoSe_2$ MLs[34,35]. This is also consistent with recent calculations which predict trion radiative recombination times longer than the neutral exciton one[36].

When the temperature increases, figure 6b shows that the trion integrated PL intensity decreases drastically and it totally vanishes at about 125 K in agreement with previous measurements[42,56]. In these previous works, the relative intensity of exciton and trion line and its dependence as a function of temperature have been analysed in the framework of a mass action law between excitons, trions and resident electrons[30,42] with electrons escaping their bound trion state due to thermal fluctuations. However the corresponding temperature dependence of the PL dynamics has not been measured yet to the best of our knowledge. Figure 6a and 6b show that the trion PL decay time decreases strongly when $T$ increases. Remarkably it varies from 15 ps at $T = 7$ K down to times smaller than 1 ps at $T > 100$ K. Considering the very different exciton and trion lifetimes measured in this study, the validity of a mass action law to explain the temperature dependence is questionable. Such interpretation based on a thermodynamical equilibrium between excitons, trions and resident electrons fails to explain the very short trion PL decay time measured at 100 K (at this temperature the exciton decay time measured in exactly the same conditions is of the order of one nanosecond, *i.e.* three orders of magnitude longer than the trion one). An alternative possible explanation could be the following: for temperatures larger than 50 K, the trion PL decay time could be governed by efficient exciton-LO phonon inelastic scattering as recently observed in $WSe_2$ with a double-Raman resonance between exciton and trion states[58]. When the temperature increases, this process is more and more efficient yielding a shorter decay of the trion luminescence intensity, which is no more driven by the trion radiative recombination time but by the trion-phonon interaction time. The fast decrease of the trion PL decay time induced by this scattering mechanism at large temperature is perfectly consistent with the drop of the trion integrated PL intensity when the temperature increases (see figure 6b). Moreover, the activation energy $E_A$ of the decay process extracted from figure 6b is $E_A = 33 \pm 2$ meV, quite close to both the trion measured binding energy $E_B(X^-) \sim 31 \pm 1$ meV (figure 1a and figure 6b inset) and optical phonon energies of IMC (in-plane relative motion of transition metal and chalcogen atoms) and OC (out-of-plane chalcogen vibration) modes ($E^{E'}_{IMC} = 35.8$ meV or $E^{A_1'}_{OC} = 30.0$ meV respectively [71,72]). Once the trion has been dissociated into an exciton and a free electron for temperatures higher than 50 K, the reverse process is very unlikely because the bimolecular trion formation time at high temperature becomes much longer than the exciton lifetime, similarly to what has been clearly evidenced in InGaAs quantum wells, where the large increase of the trion formation time with temperature has been observed[73].

The interpretation of the exciton and trion dynamics is further supported by complementary experiments performed on ML C where the density of the resident electrons can be electrically tuned. As displayed in figure 7a, the integrated luminescence intensity of the trion PL line decreases (with a simultaneous increase of the exciton line) when the applied voltage varies from -5 to 8 volts as a result of a decrease of the resident electrons in the monolayer. Figure 7b and 7c show that the exciton and trion PL decay time at 7 K do not change at all when the resident electrons density varies in this applied voltage range. This is again a strong indication that the exciton and trion population decays independently and their radiative recombination times are so short that a dynamical equilibrium between the two species cannot be established. Let us emphasize that this conclusion applies for the quasi-resonant excitation and the low temperature used here. Further investigations are required to check if this remains valid whatever the excitation energy is, *i.e.* for instance when the laser excitation energy lies above the free carriers gap[30]. Temperature dependent would also be relevant. Unfortunately the voltage dependent measurements could no be performed for larger temperatures due to a strong increase of leakage currents in the device.

## VII. Discussion and Prospects:

We have presented in the previous sections a scenario describing the time evolution of excitons from the initial non-thermal distribution following the quasi-resonant photogeneration to the thermal regime with two driving mechanisms: radiative recombination and exciton-phonon scattering.

The results and interpretations presented above are based on several assumptions which have to be discussed.

*Exciton formation process:*

Though it is now recognized that excitons physics play a crucial role on the optoelectronic properties of TMDC monolayers, the exciton formation dynamics has not been much studied. In inorganic semiconductors, two exciton formation processes are usually considered: (i) straight hot exciton photogeneration, with the simultaneous emission of LO phonons, in which the constitutive electron-hole pair is geminate; or (ii) bimolecular exciton formation which consists of the random binding of electrons and holes under the Coulomb interaction. Very few experimental results give a direct insight into the exciton formation processes. The geminate formation has been evidenced in polar semiconductors[74] whereas the second process was demonstrated in silicon, and led to the measurement of the bimolecular formation coefficient[75]. In GaAs quantum wells, several investigations also demonstrated the key role played by the bimolecular formation process[76,77,78].

In the present paper where we used quasi-resonant excitation conditions, the excitation energy lies far below the free carrier gap. Thus the exciton formation process is geminate as assumed in the previous sections.

Interestingly the measured exciton PL dynamics is very similar when the excitation energy is set below or above the free carriers gap, see for instance ref. [32,33,30,34] in MoS$_2$ or WSe$_2$ monolayers. For non-resonant excitation above the gap, the PL rise time is still very fast and no signature of bimolecular formation and energy relaxation of hot excitons can be evidenced, in contrast to III-V or II-VI quantum wells. As a consequence, we can speculate that the strong-exciton phonon coupling in TMDC monolayers yields a geminate exciton formation process whatever the excitation energy is.

*Free and localized excitons dynamics*

The TRPL experiments discussed in the previous sections have been interpreted on the basis of free exciton recombination though the PL emission lines are clearly inhomogeneously broadened (FWHM ~ 10 meV at $T$ = 4 K, see figure 1.a). In contrast the measurement of intrinsic radiative lifetime of free excitons in III-V and II-VI quantum wells could only be observed in high quality samples with narrow linewidths dominated by homogenous broadening[59,60]. For excitons in lower quality samples with Stokes shift of a few meV (energy difference between emission and absorption exciton peaks), localization and scattering processes usually yield much longer decay times of luminescence[65]. In the TMDC ML investigated in this paper, we did not measure any significant Stokes shift. This is a solid hint to believe that strong exciton localization may not occur in TMDC ML. The key difference is the much smaller size of the exciton Bohr radius (~ 0.5 nm) in TMDC ML versus ~ 10 nm in GaAs quantum wells. This has to be compared to the sample in-plane potential fluctuation sizes which are not known in 2D materials based on TMDC; it could possibly be investigated in the future with near field techniques. The small size of the exciton Bohr radius with respect to the correlation length of potential fluctuations probably explains why we measure the ultrafast intrinsic radiative lifetime in TMDC ML[79].

*Role of dark excitons*

The excitons dynamics in TMDC monolayer can depend critically of the interplay between bright and dark excitons. Basically three different kinds of excitons can be termed "dark":

(i) The spin-allowed $\Gamma$ excitons with wave-vectors $k > k_0$, which lie above the radiative window (figure 4). These dark excitons were fully considered in the previous sections.

(ii) The spin-forbidden dark excitons in the $\Gamma$ zone-centre, which lie 10-30 meV above the spin-allowed optical transitions for MoSe$_2$ ML (see dashed line in figure 4a). At low temperatures $T \lesssim 100$ K, we did not observe any interplay between these dark excitons and the bright ones which decay radiatively. For larger temperatures, a dynamical equilibrium between these bright and dark excitons could arise but the simultaneous increase of non-radiative recombination channels prevent from observing any evidence of it. We emphasize that equation (2) assumes such an equilibrium (if there is no equilibrium, $\tau_{rad}^{eff}$ must be divided by a factor 2 in equation (2)[45]).

(iii) Indirect excitons: in addition to the dark excitons (i) and (ii) which have been evidenced in many semiconductor bulk and nanostructures, the D$_{3h}$ symmetry of TMDC monolayer requires to consider a third type of dark excitons. As the carriers can populate the two non-equivalent K$_+$ and K$_-$ valley (characterized by opposite spin-orbit splitting), an exciton can be formed with a conduction band and a valence band state lying in opposite valleys[70]. The lowest indirect excitons cannot recombine radiatively since they are spin-forbidden. As a fact, at $T$ = 7 K, we did not observe any fingerprint of these dark indirect excitons. At higher temperature, higher energy spin-allowed dark excitons recombination may be possible by simultaneous phonon emission or absorption. We can speculate that they could constitute an additional reservoir of dark excitons which should again lead to the observation of longer PL luminescence component associated to the relaxation from K to $\Gamma$ valley. Crucial information such as the binding and exchange energy of such excitons are desirable in order to progress on our understanding of the role of these indirect excitons on the luminescence dynamics.

**VIII. Conclusion**

We have investigated the exciton radiative recombination dynamics in TMDC monolayers by time-resolved photoluminescence experiments with optimized time-resolution. The time-evolution of the luminescence intensity in the very first picoseconds at low temperature provides precious information on the non-thermal dynamics of excitons. Ultra-fast intrinsic radiative recombination times ($\sim 2$ ps) are measured for both $MoSe_2$ and $WSe_2$ monolayers. The temperature dependence of the luminescence decay times allows us to reveal the complex exciton dynamics controlled by the effective radiative recombination time, very efficient exciton-phonon scattering, exciton-exciton interactions as well as the interplay between the bright excitons and the various kinds of non-optically active excitons. In the future the measurement of the homogeneous linewidth with Four-Wave mixing spectroscopy in monolayers where the exciton and trion lines are well resolved should help to fully understand these complex radiative recombination processes in transition metal dichalcogenides.

*Acknowledgments*: The authors thank Jean Roux from the Hamamatsu company for the loan of a synchro-scan Streak Camera C10910 with a 0.9 ps time-resolution. We are grateful to M. Glazov for useful discussions. We thank the ANR MoS2ValleyControl and Programme Investissements d'Avenir ANR-11-IDEX-0002- 02, reference ANR-10-LABX-0037-NEXT and ERC Grant No. 306719 for financial support. X.M. also acknowledges the Institut Universitaire de France.

Figure 1

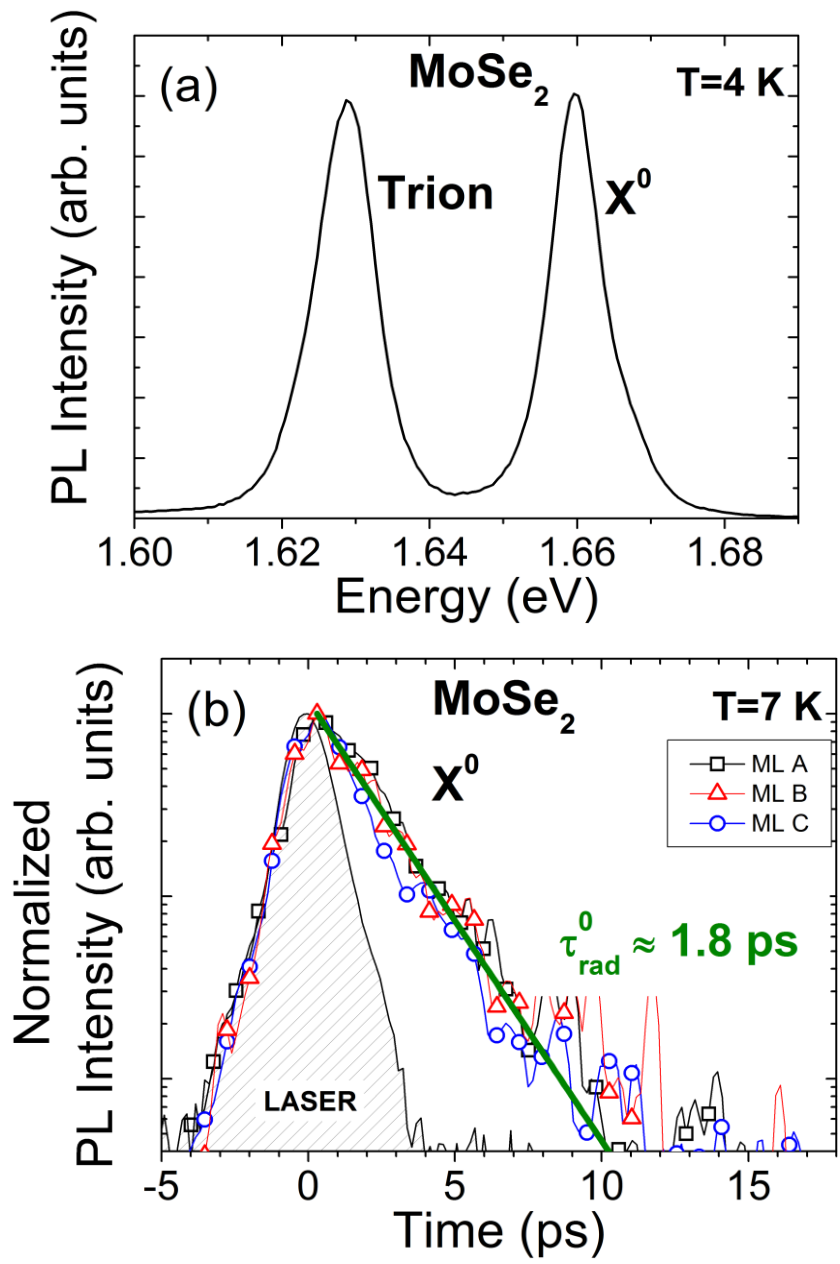

Figure 2

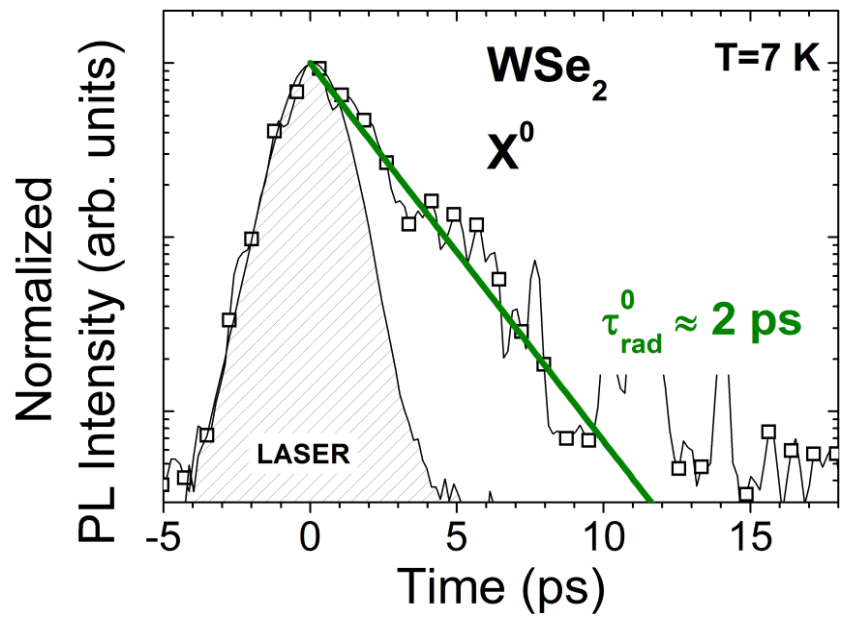

Figure 3

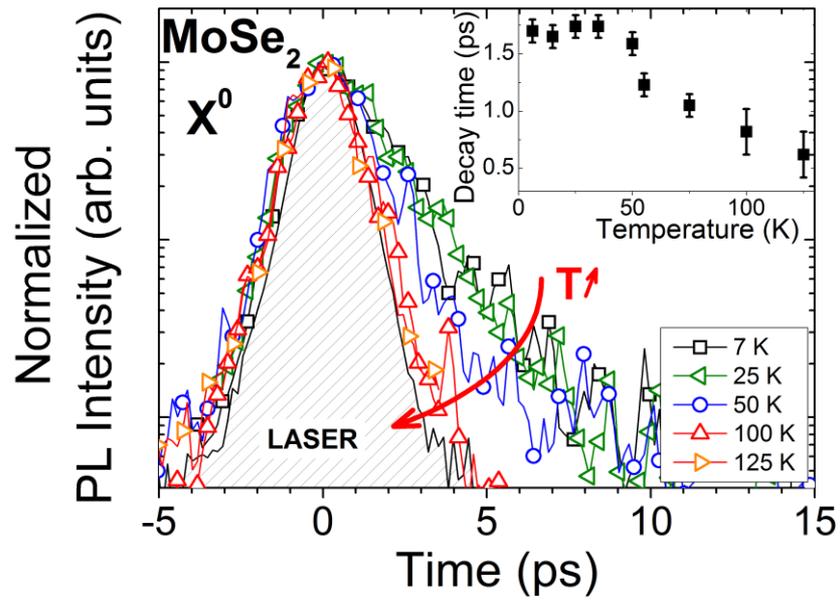

Figure 4

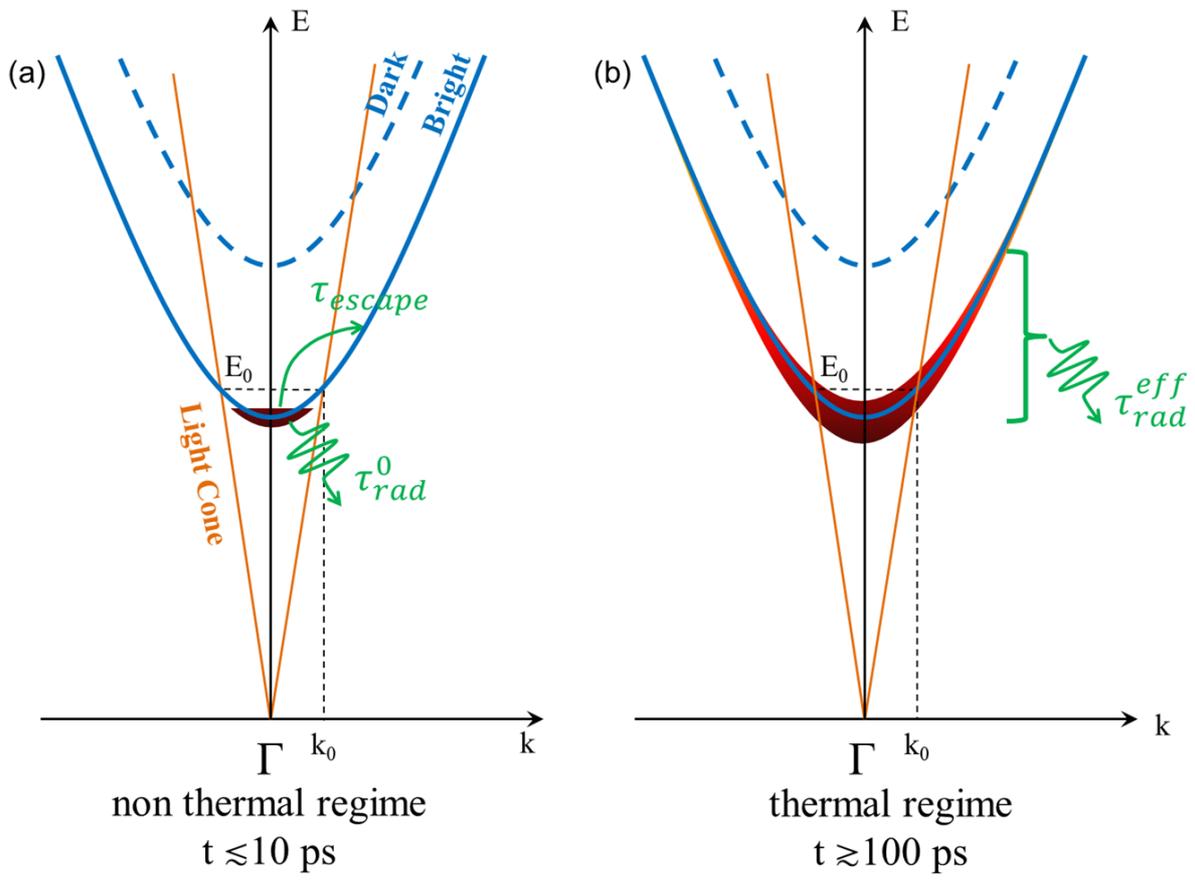

Figure 5

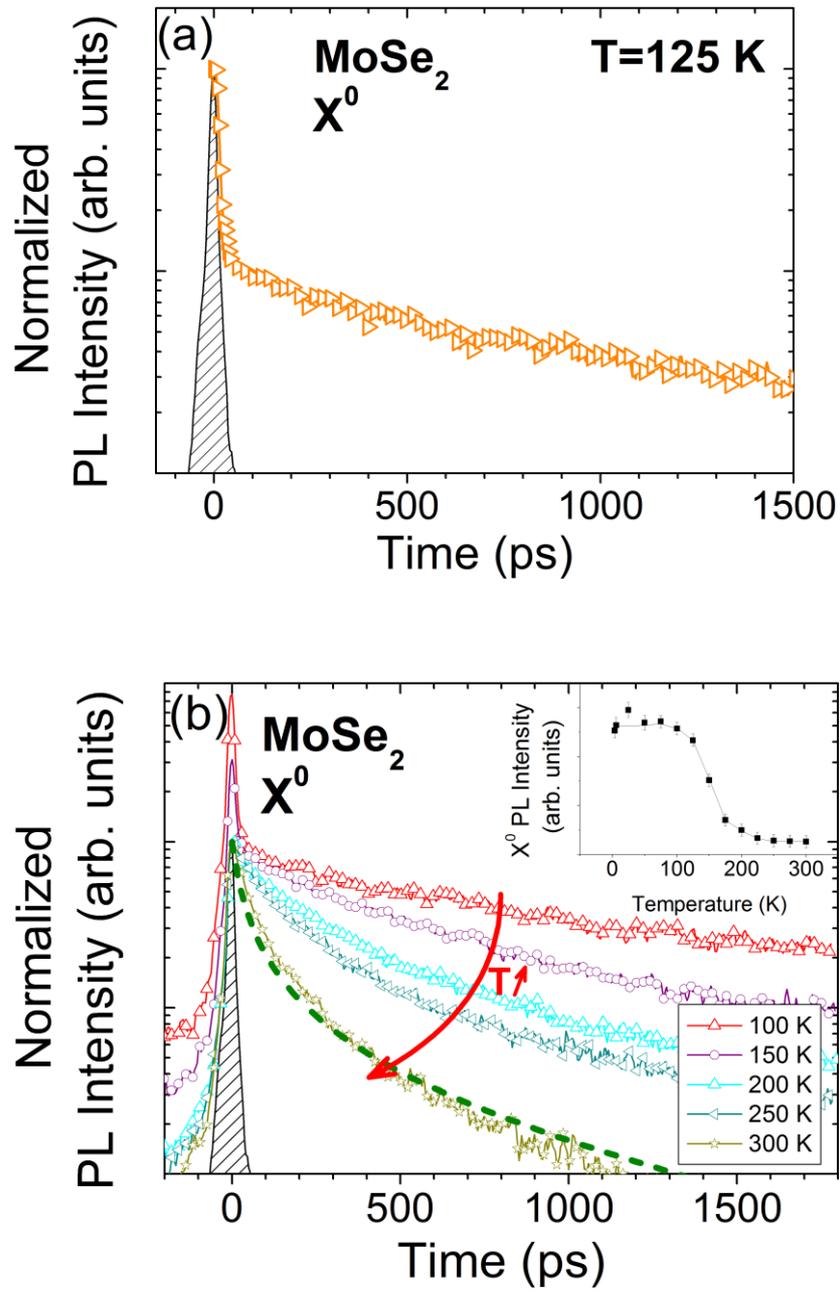

Figure 6

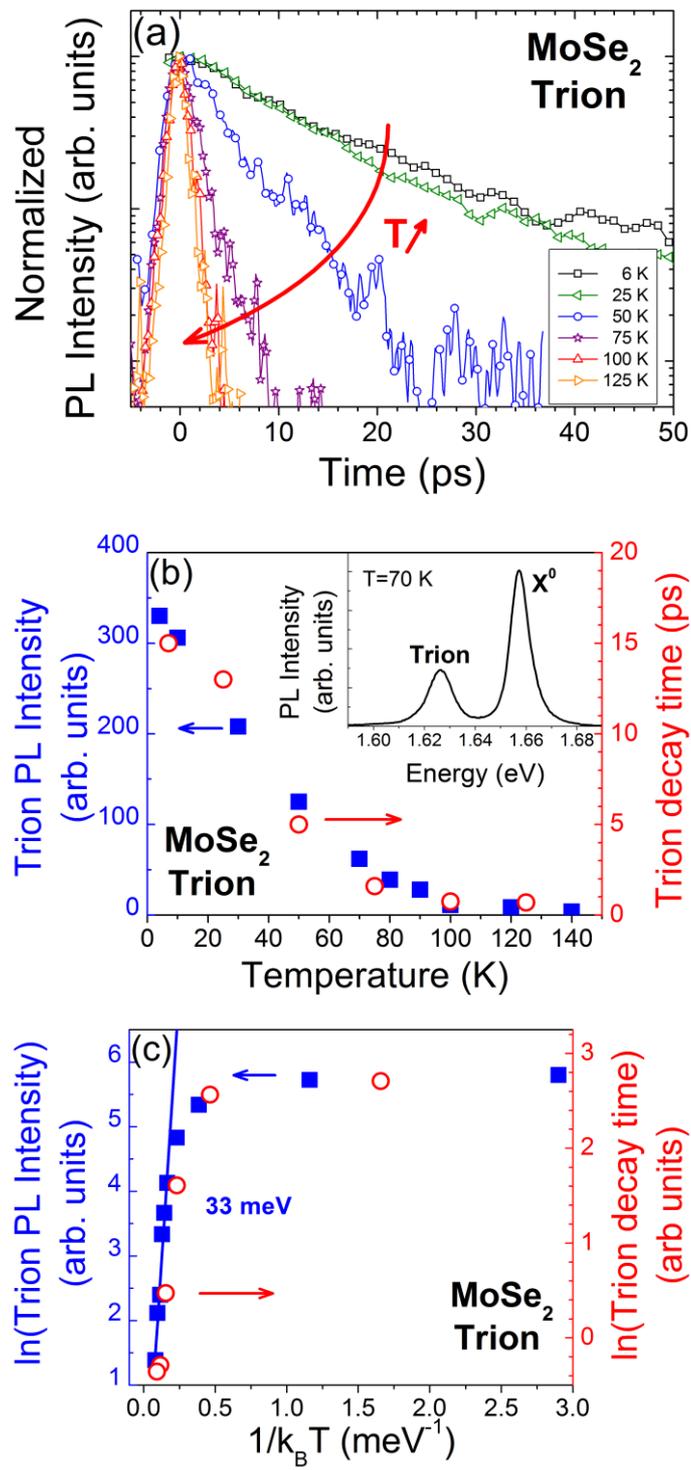

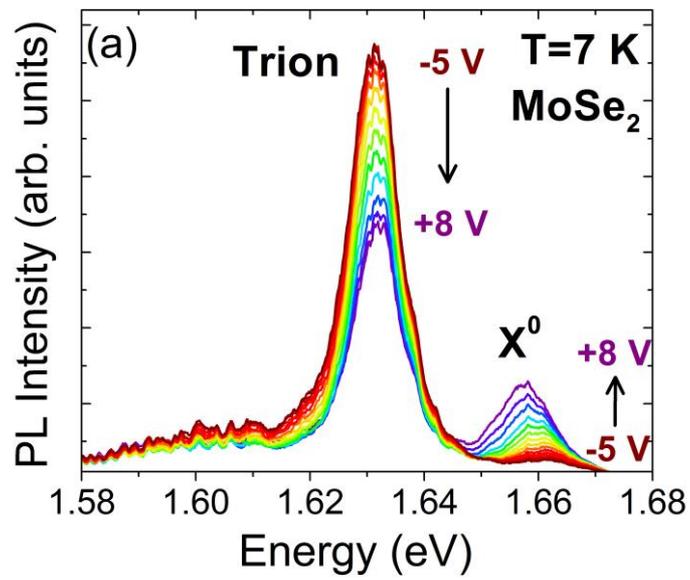

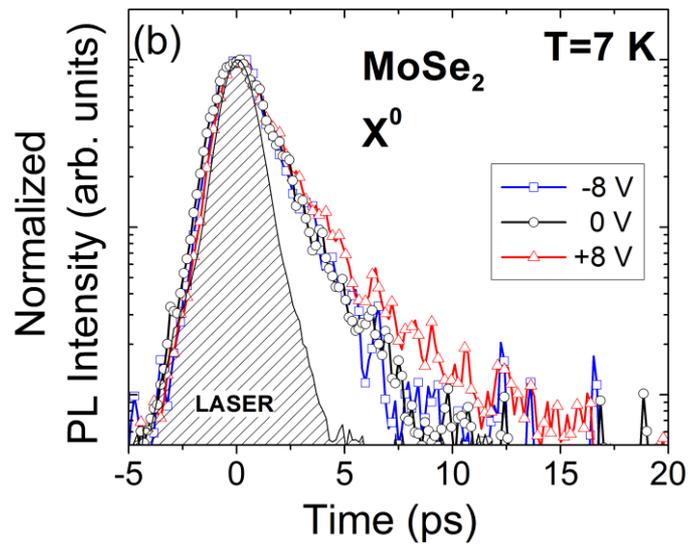

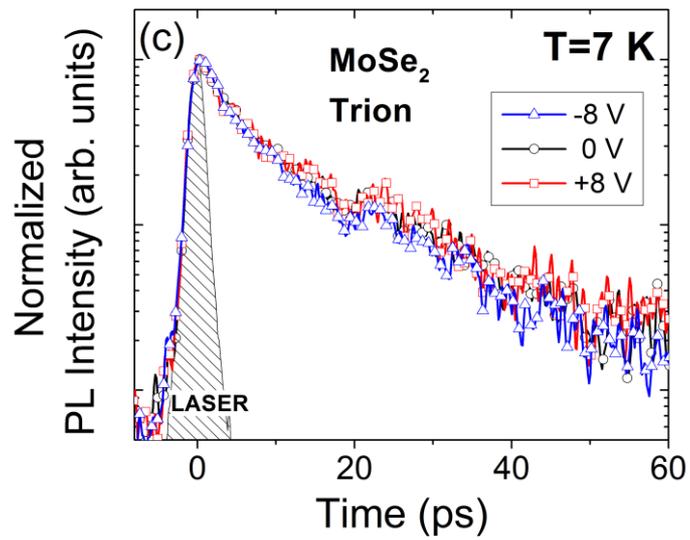

**CAPTIONS**

Figure 1. (a) Cw-photoluminescence spectrum of a MoSe$_2$ monolayer (ML A) evidencing the neutral exciton X$^0$ and the trion (charged exciton) peaks at $T$ = 4 K. (b) Time-resolved photoluminescence of three different MoSe$_2$ monolayers following quasi-resonant excitation ($E_{exc}$ = 1.746 eV) with 150 fs laser pulse ; the detection energy corresponds to the X$^0$ emission (1.66 eV). A mono-exponential fit yields $\tau_{rad}^0$ = 1.8 ± 0.2 ps. The instrument response is obtained by detecting the backscattered laser pulse (1.746 eV) on the sample surface, see the hatched area labeled « LASER ».

Figure 2. Time-resolved photoluminescence of a WSe$_2$ monolayer (ML D) following quasi-resonant excitation with 150 fs laser pulse ; the excitation and detection energy are 1.797 eV and 1.751 eV respectively. A mono-exponential fit yields 2 ± 0.2 ps.

Figure 3. Temperature dependence of the exciton photoluminescence dynamics for the MoSe$_2$ monolayer (ML A). The excitation energy is $E_{exc}$ = 1.746 eV and the detection energy is set to the peak of the X$^0$ exciton photoluminescence peak. The inset displays the photoluminescence decay time (obtained with a simple mono-exponential fit) as a function of temperature.

Figure 4. Schematics of the exciton dispersion curve in the exciton zone centre Γ for a MoSe$_2$ monolayer displaying the bright and dark excitons, as well as the radiative light cone characterized by the energy $E_0$ and the wavector $k_0$ (see text). Both the initial non-thermal regime (a) and the thermal one (b) are considered (see text); here the lattice temperature is low enough to consider a negligible population of the Γ-valley dark states.

Figure 5. (a) Exciton photoluminescence dynamics for the MoSe$_2$ monolayer (ML C, $V$ = 0 V) at $T$ = 125 K. After a fast initial drop by one order of magnitude the luminescence decays with a typical time of the order of ns. The hatched area corresponds to the instrument response time in the same conditions (the time-resolution of the Streak Camera for long kinetics is smaller than the one in figures 1-3. Consequently the amplitude of the fast initial drop of intensity in this figure is smaller than the one in figure 3 for the same temperature, due to lower time-resolution of the Streak Camera associated to the longer time-range used here).
(b) Exciton time-resolved photoluminescence from $T$ = 100 K to 300 K. The solid line is a fit of the exciton dynamics at T=300 K using equation (3) with the parameters $N_0 = 10^{11}$ cm$^{-2}$, $1/\tau_{rad}^{eff} + 1/\tau_{nr} = 1$ ns$^{-1}$ and the exciton annihilation rate $\gamma$ = 0.35 cm$^2$/s (see text). The inset presents the temperature dependence of the.

Figure 6. (a) Trion photoluminescence dynamics for a MoSe$_2$ monolayer (ML C, $V$ = 0 V) in the temperature range $T$ = 7−125 K ; above 125 K, the trion PL intensity vanishes.
(b) Temperature dependence of the cw integrated PL intensity of the trion line (blue squares) and the trion PL decay time (red circles) obtained from a mono-exponential fit of the trion kinetics measured in (a); inset : cw-photoluminescence spectrum of the MoSe$_2$ monolayer at $T$ = 70 K. (c) Same data as in (b) in log scale in order to extract the activation energy ($E_A$ ~ 33 meV).

Figure 7. (a) Cw-photoluminescence of the MoSe$_2$ monolayer (ML C) as a function of the gate voltage (the ground is connected to the back-Si contact).
(b) Exciton and trion (c) photoluminescence dynamics for three gate voltages at $T$ = 7 K.